# Manipulate Quantum Emission by Interface States between Multi-component Moiré Lattice and Metasurface

Z. N. Liu[1], X. Q. Zhao[1], Y. L. Zhao[1], S. N. Zhu[1], and H. Liu[1,*]

[1]*National Laboratory of Solid State Microstructures, School of Physics, Collaborative Innovation Center of Advanced Microstructures, Nanjing University, Nanjing 210093, China*

Email: liuhui@nju.edu.cn

In recent years, moiré lattice has become a hot topic and inspired the research upsurge of moiré lattice. In this work, we propose a method of constructing a multi-composite moiré lattice, which is composed of over three periodic component structures. Moreover, we propose the moiré lattice-metasurface structure, which can realize the multi-wavelength interface states between these kinds of moiré lattices and metasurfaces. The wavelength, polarization, and number of moiré interface states can be manipulated flexibly, with anisotropic metasurfaces. These multi-wavelength interface states are employed to enhance quantum emission (QE) and over 20 times QE efficiency can be obtained.

*Introduction*. —A moiré lattice is a composite structure formed by the overlap of two identical periodic structures[1]. Recently, in the moiré lattice of twisted double-layer graphene, it was found that at the so-called magic angle, the flat band appears near the Fermi level, and introduces nontrivial topological phases [2-4]. Mott insulating phase and superconducting phase are exhibited in the double-layer graphene [5-8].

Meanwhile, the moiré lattice brings the possibility of many other novel physical phenomena, including moiré excitons [9], fractional Chern insulators [10], natural plasmon photonic crystal[11], etc. Moiré lattices have also introduced novel interesting physical effects in artificial photonic systems [12,13]. In a two-dimensional moiré lattice, the localization-delocalization transition of light is realized experimentally [14]. The tunable topological transitions and phonon polaritons were achieved in bilayers of α-phase molybdenum trioxide (α-MoO3) [15,16]. The interlayer quantum coupling in twisted bilayer graphene may be utilized to build various atomically thin metasurfaces[17]. The formations of optical solitons are controlled by the twist angle in moiré lattices [18,19], and magic-angle lasers in nanostructured moiré lattices exhibit salient features[20]. Recently, a coupled-state theory for low-angle twisted bilayer honeycomb photonic lattices reveals a correspondence between fermionic and bosonic moiré systems [21]. Meanwhile, slow light, nonlinear effects, chiral plasmons, thermal emitters, and filters exhibit excellent properties in the moiré lattice system [17,22]. At present, the moiré lattice is formed by the superposition of two periodic structures, while the moiré lattice formed by the superposition of three or more periodic structures is rarely reported.

In this work, a design method of a multi-component moiré lattice is proposed for the first time. We combine three or more periodic structures to form a very flexible moiré lattice. Meanwhile, a new kind of multi-wavelength interface states between the moiré lattice and metal are obtained. The moiré lattice-metal structure can support multi-wavelength interface states by designing the structure of the moiré lattice.

Furthermore, we can replace the metal layer with a metasurface to manipulate the polarization of the MISs. In the experiment, this kind of interface state is used to enhance QE. An anisotropic interface state is used to control the polarization of QE.

*Design of moiré lattice-metasurface structure to realize moiré interface states.* — Firstly, let us consider a moiré lattice composed of *n* photonic crystals with different periods: $\Lambda_1, \Lambda_2, ... \Lambda_i ... \Lambda_n$. The moiré lattice can be expressed as

$$H^n(z) = \sum_{i=1}^{n} H_{\Lambda_i}(z) \delta_i(z),$$

$$H_{\Lambda_i}(z) = F_{\Lambda_i}\left(\frac{floor(\frac{z+p_i}{d}) + ceil(\frac{z+p_i}{d})}{2} d\right)$$

(1)

, where the function *ceil(x)* means the least integer $\geq x$ and the function *floor(x)* means the largest nearest $\leq x$. $F_{\Lambda_i}(z) = 1 - 2\,floor(\frac{2z}{\Lambda_i}) + 4\,floor(\frac{z}{\Lambda_i})$ is a periodic function taking the value 1 or -1, with a period $\Lambda_i$ and $\delta_i(z) = \delta(floor(\frac{z}{d}) - n \cdot floor(\frac{z}{n \cdot d}) + 1 - i)$, where $\delta(0)=1$; otherwise, $\delta=0$. $H^n(z)$ is the moiré lattice structure function. Where $H^n(z) = 1$, the refractive index equals $n_A$; otherwise, $n_B$. When $\Lambda_1 : \Lambda_2 : ... \Lambda_i : ... : \Lambda_n = q_1 : q_2 : ... q_i : ... : q_n$, where $q_1, q_2, ... q_i ... q_n$ are positive integers, the moiré lattice is periodic and the period $\Lambda$ of the moiré lattice is the least common multiple of $\Lambda_1, \Lambda_2, ... \Lambda_i ... \Lambda_n$. It is worth noting that $H_{\Lambda_i}(z)$ is a periodic function with a period $\Lambda_i$ and can characterize a photonic crystal structure. And the minimum thickness of the photonic crystals layer $d = \frac{\Lambda_1}{2q_1} = \frac{\Lambda_2}{2q_2} ... = \frac{\Lambda_i}{2q_i} ... = \frac{\Lambda_n}{2q_n}$. $p_i$ is a translation constant in the *z*-direction that can be used to tune the reflection phase at the surface of the photonic crystal. Here, we define $z=0$ as the location of the surface of the photonic crystals.

The interface between the moiré lattice and a metal layer can retain the original interface states between the original photonic crystals and metal layers. Briefly, we can design photonic crystals with different periods depending on the first band gap position. Furthermore, the phase conditions for the existence of interface states at the photonic crystal-metal interface is $\varphi_m + \varphi_{pc} = 0$, where $\varphi_m$ is the reflection phase of the metal layer surface and $\varphi_{pc}$ is the reflection phase of the photonic crystal surface[23]. And for the photonic crystal corresponding to $H_{\Lambda_i}(z)$, $\varphi_{pc}$ can be modulated by $p_i$. Thus, we can obtain different interface states as we need. In the design, we choose $n_A$=2.17 and $n_B$=1.46. Three photonic crystals defined with functions $H_{\Lambda_1}(z)$, $H_{\Lambda_2}(z)$ and $H_{\Lambda_3}(z)$ are taken as instances with $\Lambda_1 = 160$nm, $\Lambda_2 = 200$nm, $\Lambda_3 = 240$nm and $d$=20nm. When $p_1$=40nm, $p_2$=40nm and $p_3$=40nm, the first band gaps of the three photonic crystals are located near 600nm, 740nm, and 880nm, respectively. The interface states between the three photonic crystals and silver layers are obtained at these three wavelengths, respectively. Moreover, according to Eq. (1), we combine the three photonic crystals to form a moiré lattice in Fig. 1(a). In Fig. 1(c), the moiré lattice corresponding to $H^3(z)$ with period $\Lambda = 2d[q_1, q_2, q_3] = 2400$nm, inherits the band gap of the original photonic crystals (white areas) and in Fig. 1 (d), the dark blue dashed lines are the reflection spectra of the moiré lattice composed of the three photonic crystals. The reflection gaps of the original photonic crystals are also reflection gaps in the spectrum of the moiré lattice (white areas). Furthermore, at $z=0$, the phase condition for the existence of interface states of photonic crystals-silver still holds for the moiré lattice-silver in Fig. 1(b). The reflection phase of the moiré lattice has three intersection points (red pentagram) with the negative value of the silver reflection phase inside the band gaps. The wavelengths of the three intersection points are 600nm, 740nm, and 880nm. Thus, the wavelengths of the interface states of moiré lattice-silver are 600nm, 740nm, and 880nm, which is shown in the red dashed horizontal lines in Fig. 1(c) and dips in the red line in Fig. 1(d). As shown in Fig. 1(e), we demonstrate the electric field distribution of the three interface states in the moiré lattice-silver structure, which are localized at the interface between the moiré lattice and the silver layer (blue area). Here, we call the interface states between the moiré lattice and a metal layer moiré interface states (MISs).

In our moiré lattice-silver structure, the $i$-th MIS can be individually modulated by $p_i$. The moiré lattice interface is obtained by truncating the lattice and the truncated location is defined as $p_i$. With $p_i$ increasing, the truncation location of the moiré lattice is shifted to the left. The truncation location affects the reflection phase at the moiré interface. In Fig. 2, the moiré lattice corresponding to $H^3(z)$ is taken as an instance. As shown in Fig. 2(a), when $p_1$ changes from 0 to 320nm, the first MIS (dark blue line) wavelength shifts from 620nm to 570nm significantly. However, the other two MISs wavelength shifts slightly. As shown in Fig. 2(b), when $p_2$ changes from 0 to 400nm, the second MIS (green line) wavelength shifts from 770nm to 710nm significantly. However, the other two MISs wavelength shifts slightly. As shown in Fig. 2(c), when $p_3$ changes from 0 to 480nm, the third MIS (red line) wavelength shifts from 926nm to

847nm significantly. However, the other two MISs wavelength shifts slightly. The above results show that $p_i$ can be used to individually modulate the $i$-th MIS wavelength. We select the moiré structures (red, green, and blue dots) in Fig. 2 (a)(b)(c) corresponding to different $p_1$, $p_2$, and $p_3$ and plot the reflection phase (red, green, and dark blue lines) in Fig. 2 (d)(e)(f). $p_i$ only has a significant effect on the reflection phase modulation of the moiré lattice in the corresponding white region, but a small effect on other regions. This leads to a shift in the wavelength of the $i$-th MIS corresponding to $p_i$ in the moiré lattice-silver structure, as shown in Fig. 2(g)(h)(i). In Fig. 2(g), with $p_1$ increased, the first MIS (dips in white areas) blueshifts but the other two MISs barely move. In Fig. 2(h), with $p_2$ increased, the second MIS (dips in white areas) blueshifts but the other two MISs barely move. In Fig. 2(i), with $p_3$ increased, the third MIS (dips in white areas) blueshifts but the other two MISs barely move. Thus, $p_i$ can be used to individually tune the $i$-th MIS wavelength by controlling the reflection phase of the moiré lattice.

Besides, for the moiré lattices composed of more photonic crystals, the combination rule of the multi-wavelength MISs given above is still valid. For instance, we add $H_{\Lambda_4}(z)$ with $\Lambda_4 = 280$nm to $H^3(z)$ to obtain $H^4(z)$ in Fig. 1(f) and when $p_4$=300nm, the photonic crystal corresponding to $H_{\Lambda_4}(z)$ has interface state with silver at 1040nm wavelength. In Fig. 1(g), the moiré lattice corresponding to $H^4(z)$ with period $2d[q_1, q_2, q_3, q_4]$ and $d$=20nm, inherits the photonic band gap of the original photonic crystals and the interface states are also retained. As shown in Fig. 1(h), we realize quadruple MISs in the moiré lattice-silver structure. According to Eq. (1), we can achieve more MISs in a moiré lattice-silver structure if more photonic crystals are added.

*Moiré interface states for enhancement of quantum emission.* —In the above discussion, we find MISs can be obtained at the moiré lattice-silver interface. We can use the multi-wavelength interface states to increase the QE efficiency if one MIS is resonant at the pumping wavelength, and the other MIS is resonant at the QE wavelength. Here, we utilized a pump laser with wavelength at 450nm to excite QE with wavelength at 610nm. For the photonic crystal, we choose tantalum pentoxide as material A (the refractive index $n_A$=2.17) and silicon dioxide as material B (the refractive index $n_B$ =1.46). The moiré lattice is designed based on Eq. (1) composed of two periodic components. We choose the parameters $\Lambda_1 = 120\text{nm}$, $\Lambda_2 = 160\text{nm}$ and get $d$=20nm, $p_1$=60nm, and $p_2$=20nm, which ensure there are two MISs of moiré lattice-silver with wavelength at 450nm(the first MIS) and 610nm(the second MIS). The moiré lattice corresponding to $H^2(z)$ is shown in Fig. 3(a). To increase the QE efficiency by moiré lattice, we added a polymethylmethacrylate (PMMA) layer mixed with rare earth (Europium), which can be excited by a laser with wavelength shorter than 610nm and emits fluorescence at 610nm, on top of the moiré lattice, marked as the red layer in Fig. 3(a). The thickness and the refractive index of the PMMA layer are 160nm and 1.62. And interface states can be obtained at the interface between the moiré lattice and a silver layer. In Fig. 3(b), the red line is the reflection spectrum of the moiré lattice-silver structure and the dark blue dashed line is the reflection spectrum of the moiré lattice. The two MISs correspond to the two dips (wavelength at 450nm and 610nm) of the red line. This double wavelength MISs between the silver layer and the moiré lattice can enhance the pump laser and QE simultaneously, thus increasing the QE efficiency.

To illustrate the enhancement of QE efficiency in the moiré lattice-silver structure, we use a temporal coupled state theory(see Appendix)[24]. As shown in Fig. 3(c), we give the calculated $\Gamma$ (QE intensity) at different pumping wavelengths, which are shown as dots in Fig. 3(b). Clearly, the QE intensity is greatest at the pumping wavelength of 450nm. As the pumping wavelength is shifted away from the MIS wavelength, the QE intensity is decreased due to the pumping laser departing from the resonance state. The calculation results show that the QE intensity can be dramatically enhanced with MISs. Experimentally, we use a tunable femtosecond laser to excite a quantum emitter (rare earth) in the moiré lattice-silver interface. The pump laser is input on the silver side and the QE intensity is measured on the silver side. As shown in Fig. 3(d), we change the laser wavelength from 440nm to 470nm spaced by 5nm, keep the power constant, and measure the QE intensity. The QE intensity gets the maximum value at the laser wavelength of 450nm. The inset shows the peak of QE intensity with different pumping wavelengths, which is in good agreement with the theoretical calculations.

Besides, we compare the QE intensity produced in three different structures, a glass substrate, a Tamm structure, and the moiré lattice-silver structure we mentioned above (corresponding to $H^2(z)$), whose electrical field profiles are given in Fig.4(a-c). Here, Tamm structure is defined as a periodic photonic crystal and metal layer, which has a single wavelength resonance interface state at 450nm. Usually, Tamm structure shows sharp narrow peak resonance in reflection and transmission spectra. Utilizing this property, optical absorbers, filters, lasers, sensors, etc. can be fabricated. However, the moiré lattice-silver structure not only possesses the resonance properties of Tamm structures but also has no limit on the number of resonance states. No field is localized in a substrate without period in Fig. 4(a). Only one electric field of wavelength 450nm is localized in Tamm structure in Fig. 4(b). It is worth noting that since the moiré lattice supports the resonance interface states at both two wavelength 450nm and 610nm. As shown in Fig. 4(c), in the moiré lattice-silver structure, both the electric fields of two wavelengths are localized at the interface simultaneously. These different electric field distributions affect the coupling coefficient which affects the QE intensity. To analyze the effect of different structures on QE intensity, we obtained the coupling coefficients in calculations, $|\kappa_2| = 6.62 \times 2\pi$ *THz* for the moiré lattice-silver structure, $|\kappa_1| = 1.26 \times 2\pi$ *THz* for the single resonant Tamm structure, and $|\kappa_0| = 0.43 \times 2\pi$ *THz* for the glass substrate. Since $|\kappa_2| > |\kappa_1| > |\kappa_0|$, the moiré lattice-silver structure has the best QE efficiency. In ref. [24], the two resonant states are not located at the same locations and are separated by a metal layer. In the structure we can obtain both the resonant states at the same interface between the metal and the moiré lattice, which makes the QE intensity stronger. Experimentally, we use a femtosecond laser with a wavelength of 450 nm to excite the quantum emitters in the three structures with the same incident power. The measured QE intensities are compared in Fig. 4(d). The best QE intensity is produced in the moiré lattice-silver structure, which is 5 times of the Tamm structure and 20 times of the glass substrate.

*Metasurface modulated moiré interface states.* — Metasurfaces are two-dimensional artificial nanostructures consisting of periodic subwavelength unit-cells. In recent years, the rapid development of metasurfaces bring us many interesting applications, such as beam-steering, surface plasmon polariton coupling, metalenses, meta-waveplates, ultrasensitive sensing, biosensors, and dynamical metasurfaces[25,26]. In applications, we expect more flexibility in tuning the MISs, which can be achieved by replacing the metal layer with a metasurface. With the help of the metasurface, the MISs can be manipulated by engineering the reflection phase of the metasurface. With an anisotropic metasurface, we can manipulate the polarization of the MISs. The moiré lattice-metasurface structure is shown schematically in Fig. 5(a). The red area indicates the PMMA layer(150nm) and the blue part indicates the anisotropic metasurface composed of subwavelength nanoslits. The nanoslits are periodical along the X direction with a thickness of 35nm and a period P=150nm. The width of a single slit is $d_m$, which is a changed parameter in the designing. We fabricate four samples with different slit widths $d_m$= 0 (silver layer), 50, 75, and 100nm. The calculated negative value of the reflection phases of X polarized light and Y polarized light by the metasurfaces are shown in Fig. 5(b), respectively. For reference, the calculated reflection phase of the moiré lattice is also shown in Fig. 5(b) as a black solid line. In Fig. 5(b), the dashed lines are the reflection phase of X polarized light, the solid line is the reflection phase of Y polarized light and the different color lines correspond to the different slit widths of the metasurfaces. The different polarized waves show different changing trends with increasing $d_m$. As $d_m$ increases, the reflection phase of the X polarized light by the metasurfaces shifts upward but the Y polarized light by the metasurfaces shifts downward. As a result, the intersection points of the reflection phase of the moiré lattice with the negative value of the reflection phases of the different polarized waves by the metasurfaces shift in the opposite direction. This makes the wavelengths of the two different polarized MISs shift in opposite directions. As shown in Fig. 5(c, d), we calculated the reflection spectra of the different polarized waves by moiré lattice-metasurface structures in simulation, and for reference, the reflection spectrum of moiré lattice (orange line) also is shown. In Fig. 5(c), the wavelength of the X polarized MISs (dips in white areas) shifts to a shorter wavelength and then disappear outside the band gap with increasing $d_m$. However, in Fig. 5(d), the wavelengths of the Y polarized MISs (dips in white areas) shift to longer wavelengths with increasing $d_m$. The experimentally measured reflection spectra of different polarized states of the moiré lattice-metasurface structure are shown in Fig. 5 (e) and (f), which are in good agreement with the simulation results Fig. 5 (c) and (d), respectively. The above results show that by tuning the metasurface structural parameter $d_m$, we can flexibly manipulate the resonance wavelength of the different polarized states of MISs.

The above results show that the anisotropic metasurfaces can be used to control the polarization of the interface states. As a result, we can apply the anisotropic metasurfaces to control the polarized states of QE. Here, we still used temporal coupled state theory to calculate the QE intensities with different polarized states in Fig. 6(a). In Fig. 6(a), the red line is the X polarized QE intensity with $d_m$ =50nm and the pumping wavelength is 445nm. The blue line is the Y polarized QE intensity with $d_m$ =50nm and the pumping wavelength is 450nm. It can be seen that the Y polarized QE intensity is nearly three times of the X polarized QE intensity. This is because both the Y polarized pumping (blue dashed vertical line) and QE wavelengths (red dashed vertical line) are at the MISs wavelength positions, as shown in Fig. 5(d). However, the X polarized QE wavelength (red dashed vertical line) is not at the MIS wavelength position, as shown in Fig. 5(c). In Fig. 6(b), we use a 450nm femtosecond laser to excite a quantum emitter in the moiré lattice-metasurface structure and then the Y polarized QE intensity is measured on the metasurface side (blue line). Then, we use a 445nm femtosecond laser with the same incident power to excite a quantum emitter in the moiré lattice-metasurface structure, and then the X polarized QE intensity is measured on the metasurface side (red dashed line). The experimentally measured results in Fig. 6(b) are in good agreement with the theory in Fig. 6(a). The above results show the metasurfaces can modulate not only the polarized state of QE but also the QE intensity. In Fig. 6 (c), we used temporal coupled state theory to calculate the Y polarized QE intensity in the different moiré lattice-metasurface structures with slit widths ($d_m$= 0, 50, 75, 100nm). The pumping wavelengths are shown in the black, red, green, and dark blue dots (left side white area) for different moiré lattice-metasurface structures with slit widths ($d_m$= 0, 50, 75, 100nm) in Fig. 5(d), respectively. The QE wavelength positions are shown in the red dashed line and the reflectivity (QE wavelength) corresponding moiré lattice-metasurface structures ($d_m$ = 0, 50, 75, 100nm) are shown as the black, red, green, and dark blue dots in the second gap in Fig. 5(d), respectively. In Fig. 6(c), the comparison between the QE intensity of different samples is decreased from the best to the worst as $d_m$=50nm, 75nm, 100nm, and 0nm. This is because the change of $d_m$ shifts the wavelength of the MISs to depart away from the QE wavelength(610nm). In Fig. 6(d), we use a tunable femtosecond laser to excite a quantum emitter in the moiré lattice-metasurface structures to produce QE. For different samples, we choose different pumping wavelengths to excite the quantum emitter, which is shown as the black, red, green, and dark blue dots on the left in Fig. 5(f). Keeping the incident power unchanged, we measure the Y polarized QE intensity of different moiré lattice-metasurface structures. The experimentally measured results in Fig. 6(d) are in good agreement with the theoretical calculations in Fig. 6(c). The above results show that by tuning the metasurface, we can manipulate the polarization state and intensity of the QE.

*Conclusion and discussion.* —In conclusion, we propose a method of constructing a multi-component moiré lattice. The moiré lattice-metasurface structure can support multi-wavelength MISs, of which the wavelength, polarization, and number can be manipulated flexibly. Furthermore, based on the MISs we propose a feasible method to realize the metasurface modulate the polarized state of QE, which can be generalized to nonlinear optics or other light-matter interaction processes with multi-wavelength waves. We believe that the designing method of constructing multi-component moiré lattice and the MISs can be extended to other kinds of moiré physical systems in the future. Some special other interesting applications are possible to be obtained, such as optical detection devices, filters, and sensors.

Acknowledgements

This work was financially supported by the National Natural Science Foundation of China (Nos. 92163216, 92150302, and 62288101).

*Corresponding author.

liuhui@nju.edu.cn

**APPENDIX: TEMPORAL COUPLED MODE THEORY FOR MANIPULATING QUANTUM EMISSION BY MOIRE INTERFACE STATES**

To explain the moiré interface state (MIS) structure enhanced quantum emission efficiency, we used temporal coupled mode theory. In this theory, first MIS and second MIS respectively refer to cavity modes 1 and 2, which are described as a kind of resonance with amplitude $A = [a, b]^T$. Among them, the first MIS corresponds to the higher frequency mode, which refers to the pump laser, and the second MIS corresponds to the lower frequency mode, which is the emission wavelength of quantum emitters. Due to the quantum emitter with a two-energy system, it enables the coupling of two orthogonal cavity modes with different frequencies. Besides, the internal leap process of quantum emitters is unidirectional, so it can be excited by photons of higher frequencies and converted into photons of lower frequencies, but the inverse condition is not allowed. Therefore, we deem that the coupling of the first MIS and the second MIS cavities is unidirectional, that is, the first MIS cavity can couple to the second MIS cavity, while the second MIS cavity cannot couple in turn. The first MIS cavity in our structure is driven by an applied pump laser, while the second MIS cavity is excited by coupling between the two cavities through quantum emitters. And

the coupling equations are:

$$\frac{da}{dt} = (j\omega_a - \gamma_1)a + S_+,$$
$$\frac{db}{dt} = (j\omega_b - \gamma_2)b + \kappa a \quad (A1)$$

where $\omega_a$ ($\omega_b$) is the eigenfrequency of the first MIS (second MIS). $S_+$ is the intensity of interaction between the first MIS cavity and the outside environment. $\gamma_1$ ($\gamma_2$) is the attenuation rate from the radiative damping and can be expressed in terms of the full width at half maximum of resonant wavelength. $\kappa$ represents the coupling coefficient between the two cavities, which is directly proportional to the overlapping integral of the electric field:

$$\kappa = \frac{(\varepsilon_0 - \varepsilon_{PMMA})\omega_0 \int_{PMMA} d\mathbf{r}^3 \mathbf{E}_1^* \cdot \mathbf{E}_2}{\langle \phi_1 | \phi_1 \rangle},$$
$$\langle \phi_{1(2)} | \phi_{1(2)} \rangle = \sum_{m \in all} \int_m d\mathbf{r}^3 (\varepsilon_m \mathbf{E}_{1(2)}^* \cdot \mathbf{E}_{1(2)} + \mu_m \mathbf{H}_{1(2)}^* \cdot \mathbf{H}_{1(2)}) \quad (A2)$$

where $\varepsilon_0$ and $\varepsilon_m$ are the absolute permittivity for the vacuum and the materials, respectively. $\mu_0$ and $\mu_m$ are respectively the permeability of the vacuum and the materials. $\mathbf{E}_{1(2)}$ and $\mathbf{H}_{1(2)}$ are respectively the electric and magnetic field distribution of resonant modes. Here, we use the normalized electric field i.e., $\langle \phi_1 | \phi_1 \rangle = \langle \phi_2 | \phi_2 \rangle$.

In our structure, the second MIS cavity is coupled with the wavelength of the quantum emitter embedded in a moiré lattice-metal configuration. The leakage of the second MIS cavity is expressed as $\Gamma = |\gamma_2 b|$. To simplify the model, we consider the interaction strength of the first MIS cavity with the pump laser as constant 1, then we

can obtain the emission amplitude: $\Gamma = \dfrac{\kappa\gamma_2 S_+}{\sqrt{\left[(\omega_l-\omega_a)^2+\gamma_1^2\right]\left[(\omega-\omega_b)^2+\gamma_2^2\right]}}$, where $\omega_l$ indicates the frequency of the pump laser. Clearly, the quantum emission intensity $\Gamma$ is affected by the pump laser frequency $\omega_l$ and the coupling strength $\kappa$. As shown in Fig. 3(c), we calculated the quantum emission intensity $\Gamma$ at different frequencies of the pump laser $\omega_l$. The inset shows the peak intensity of quantum emission at different laser wavelengths. Clearly, the quantum emission intensity is greatest at the pump laser wavelength of 450nm and coupling coefficient $|\kappa_2| = 6.62\times 2\pi$ *THz*. The calculation results show that the quantum emission intensity can be greatly improved in the moiré interface state structure.

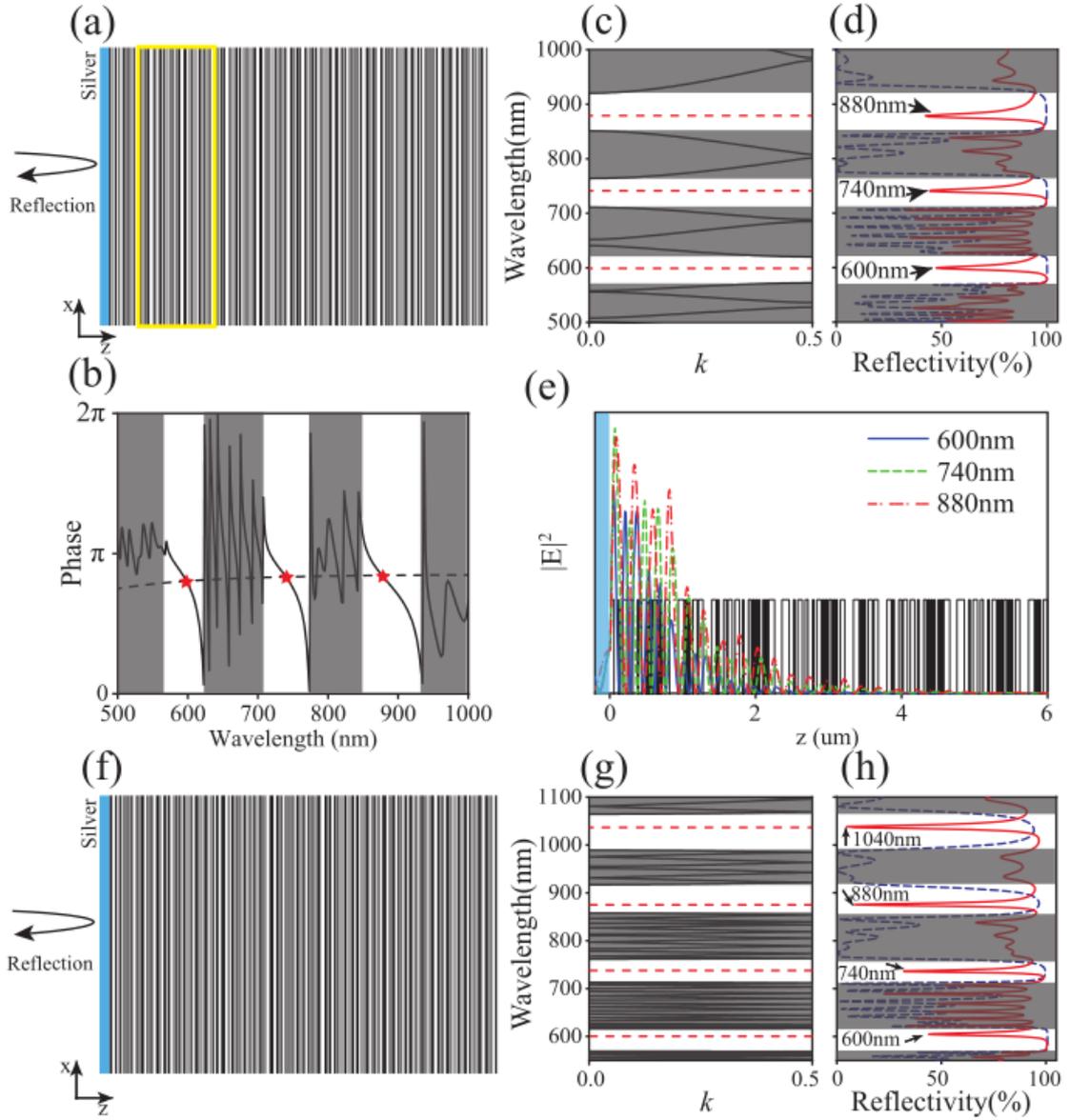

**Figure 1** (a) the moiré lattice-silver structure (the moiré lattice corresponding to $H^3(z)$). The yellow box is a moiré cell. The curved arrow on the left side is a reflection schematic. (b) The dashed black line represents the calculated negative value of the reflection phase of the silver layer. The black line shows the reflection phase of the moiré lattice. The location of the intersection (red pentagram) of the two lines is MISs. (c) show the band of the moiré lattice and the white regions is the band gaps with the MISs. The red horizontal dashed line is the MISs. (d) The dashed blue line is the reflection spectrum of the moiré lattice and the red line is the reflection spectrum of the moiré lattice-silver structure. (e) are electric field distribution of moiré interface states of 600nm, 740nm, and 880nm, respectively. The blue area is a silver layer. (f) show moiré lattice- silver structure (the moiré lattice corresponding to $H^4(z)$ ). The curved arrow on the left side is a reflection schematic (g) show the band of the moiré lattice corresponding to $H^4(z)$ and the white regions are the band gaps. The horizontal dashed red line is the MISs. (h) the dashed blue line is the reflection spectrum of the moiré lattice and the red line is the reflection spectrum of the moiré lattice-silver structure.

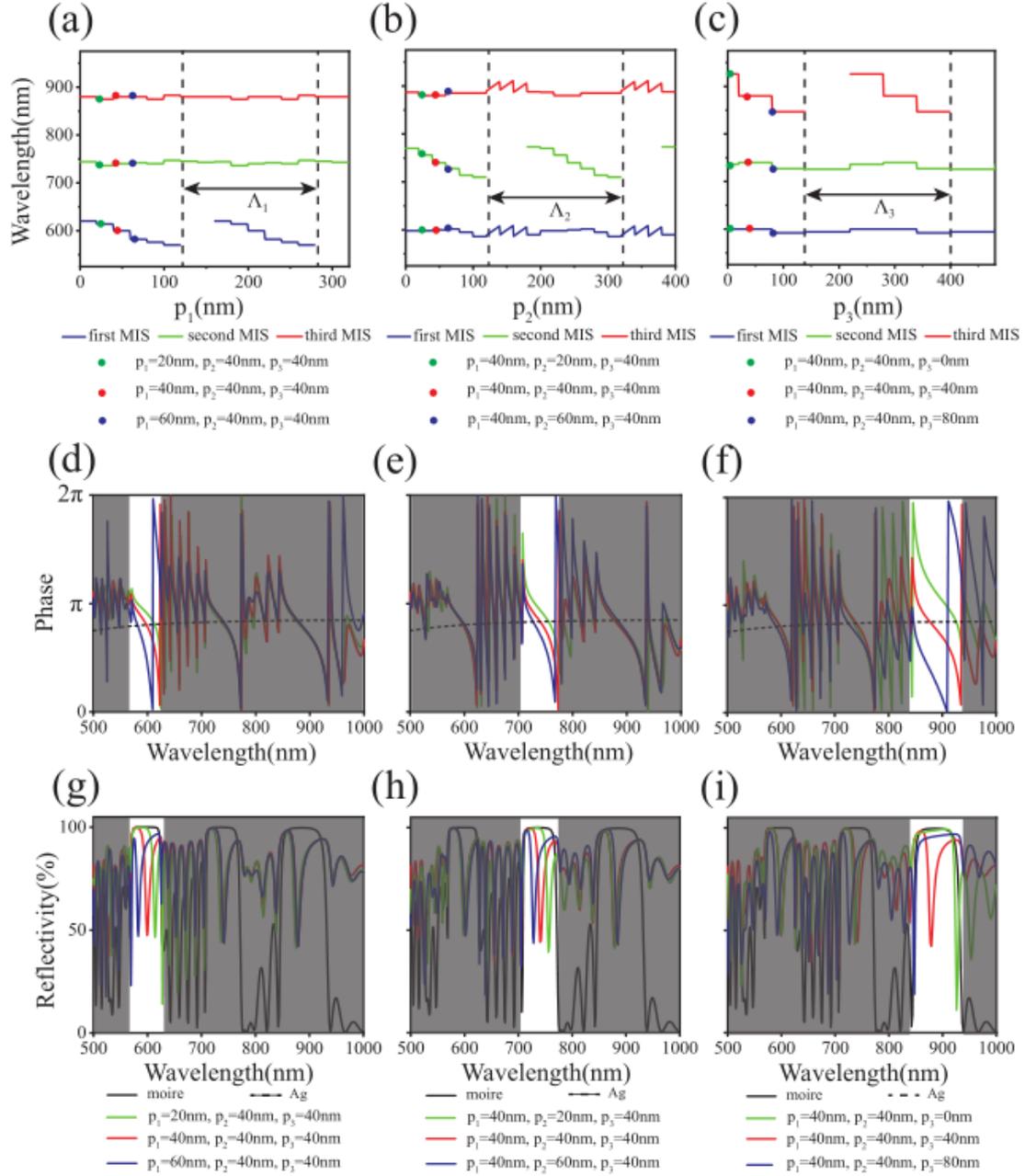

**Figure 2** (a) (b) (c) shows the change in the wavelength of the MISs as $p_1$, $p_2$, and $p_3$ change, respectively. (d)(e)(f) shows the red, green, and dark blue lines are the reflection phase of moiré lattice for different $p_1$, $p_2$ and $p_3$ cases, respectively and the black dash line is the negative of reflection phase of the silver layer. (g)(h)(i) is the reflection spectrum of the corresponding moiré lattice-silver structure. The black line is the reflection spectrum of the moiré lattice corresponding to $H^3(z)$.

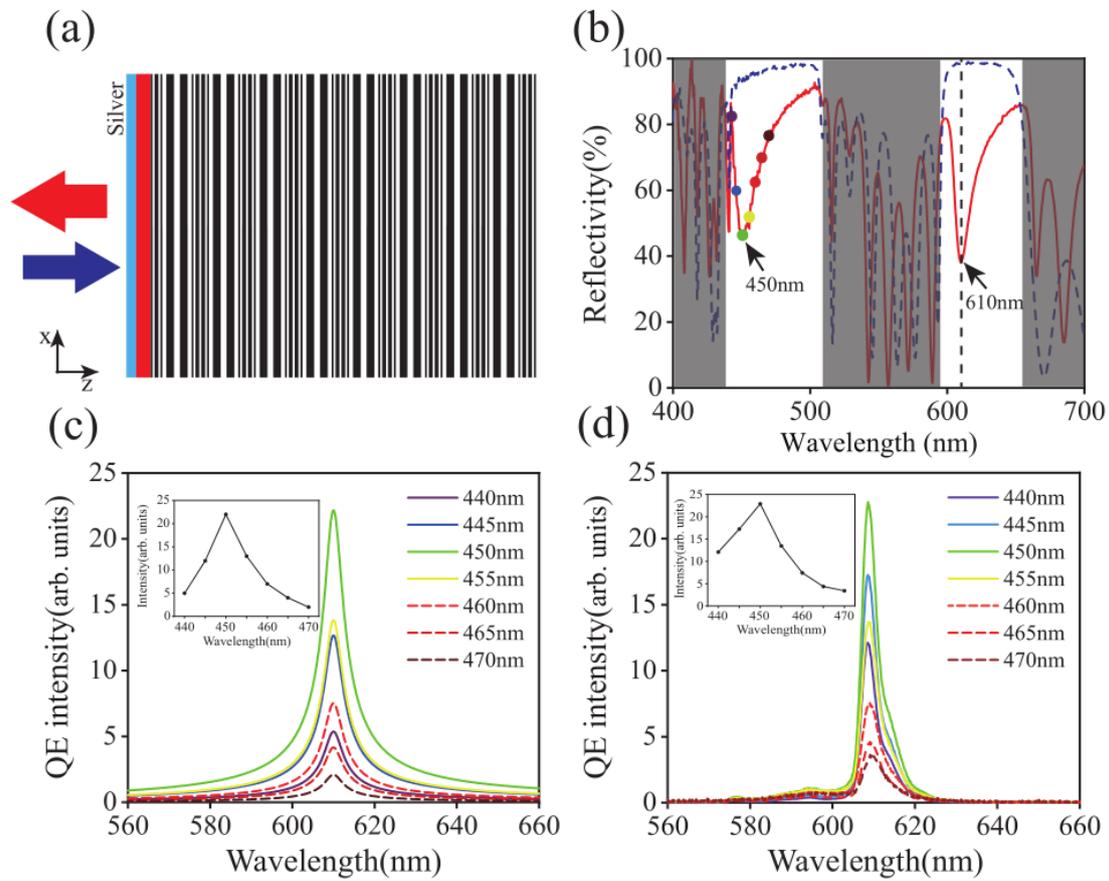

**Figure 3** (a) Schematic diagram of quantum excitation. The dark blue arrow indicates pump light incident from the silver layer, red arrows indicate QE from the silver layer side. (b) the dotted blue line (red line) is the experimental reflection spectrum of the moiré lattice (moiré lattice-silver structure). The dots correspond to the different pumping laser wavelengths in (c) and (d). (c) The QE intensity with pumping wavelength. (d) The experimental measured QE intensity with pumping wavelength.

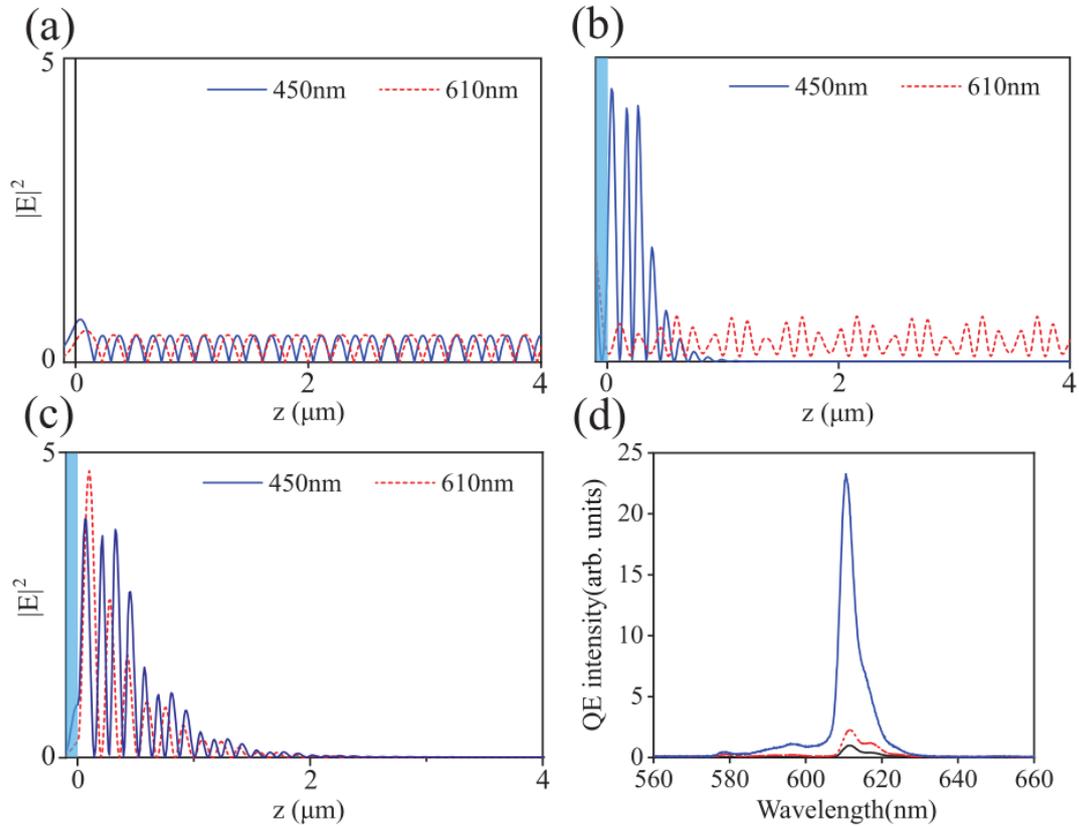

**Figure 4** (a) are the electric field distribution of 450nm(blue line), and 610nm(dashed red line) of a glass substrate, respectively. The black line is the interface between the glass substrate and the air. (b) are Tamm mode distribution of 450nm (blue line) and electric field distribution of 610nm (dashed red line). The blue area is silver. (c) are moiré interface modes electric field distribution of 450nm (blue line), and 610nm (dashed red line), respectively. The blue area is silver. (d)The blue line is the QE intensity of the moiré lattice-silver structure, the dashed red line is the QE intensity of the original photonic crystal-sliver structure, and the black line is the QE intensity of the glass substrate.

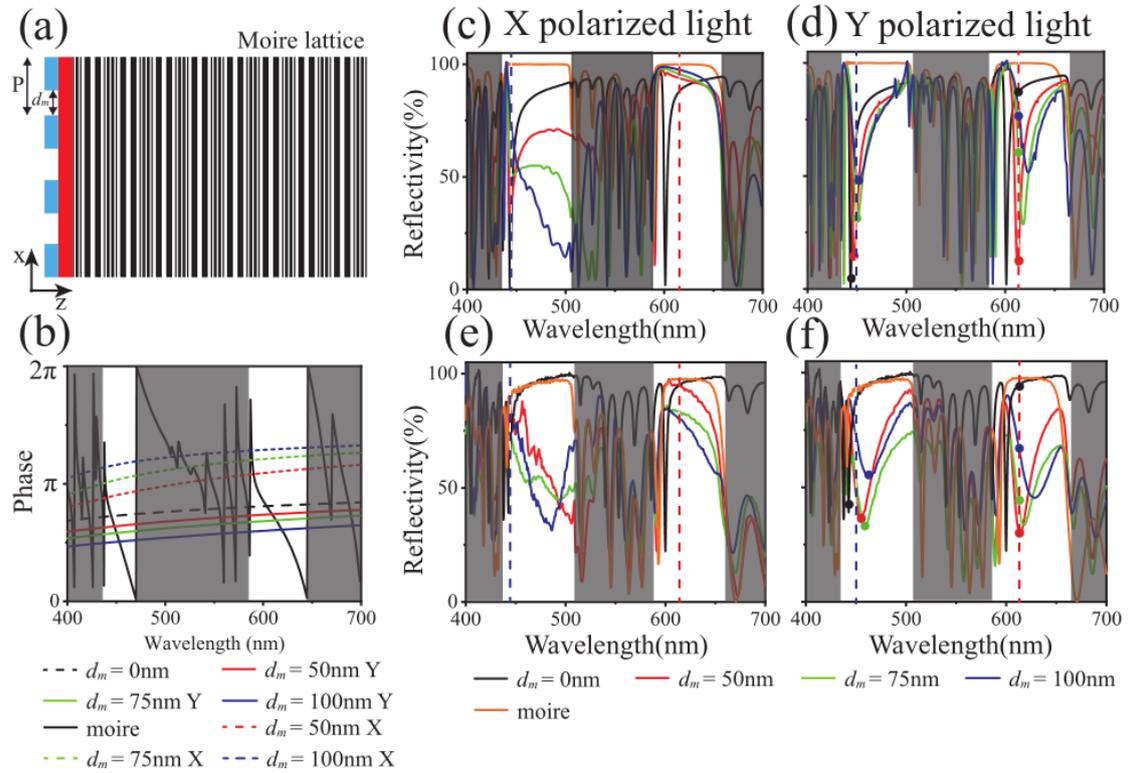

**Figure 5** (a) the moiré lattice-metasurface structure which supports MISs. (b)The solid lines are the calculated negative value of the reflection phase of Y polarized light by the metasurfaces and the dashed lines are the calculated negative value of the reflection phase of X polarized light by the metasurfaces. The black, red, green, and dark blue lines represent slit widths $d_m$ = 0 (equivalent to the silver layer), 50, 75, and 100nm of the metasurfaces, respectively. The black line shows the reflection phase of light by the moiré lattice. The reflection spectrums of the moiré lattice-metasurface structure are shown in (d) and (f) the Y polarized light and (c) and (e) the X polarized light, where c, d, and (e, f) are the numerical (experimental) results. The black, red, green, and dark blue lines represent slit widths $d_m$=0 (silver layer), 50, 75, and 100nm, respectively. The orange line represents the reflection spectrum of the moiré lattice. (c)-(f)The white regions are the band gaps we care about. The dashed red lines on the right side white region are QE wavelength and the dashed blue lines on the left side white region are selected pumping wavelength in Fig. 6(a). (d)(f) The dots on the right side white region are leaks of QE for different structures and the dots on the left side white region are selected pumping wavelength in Fig. 6(c).

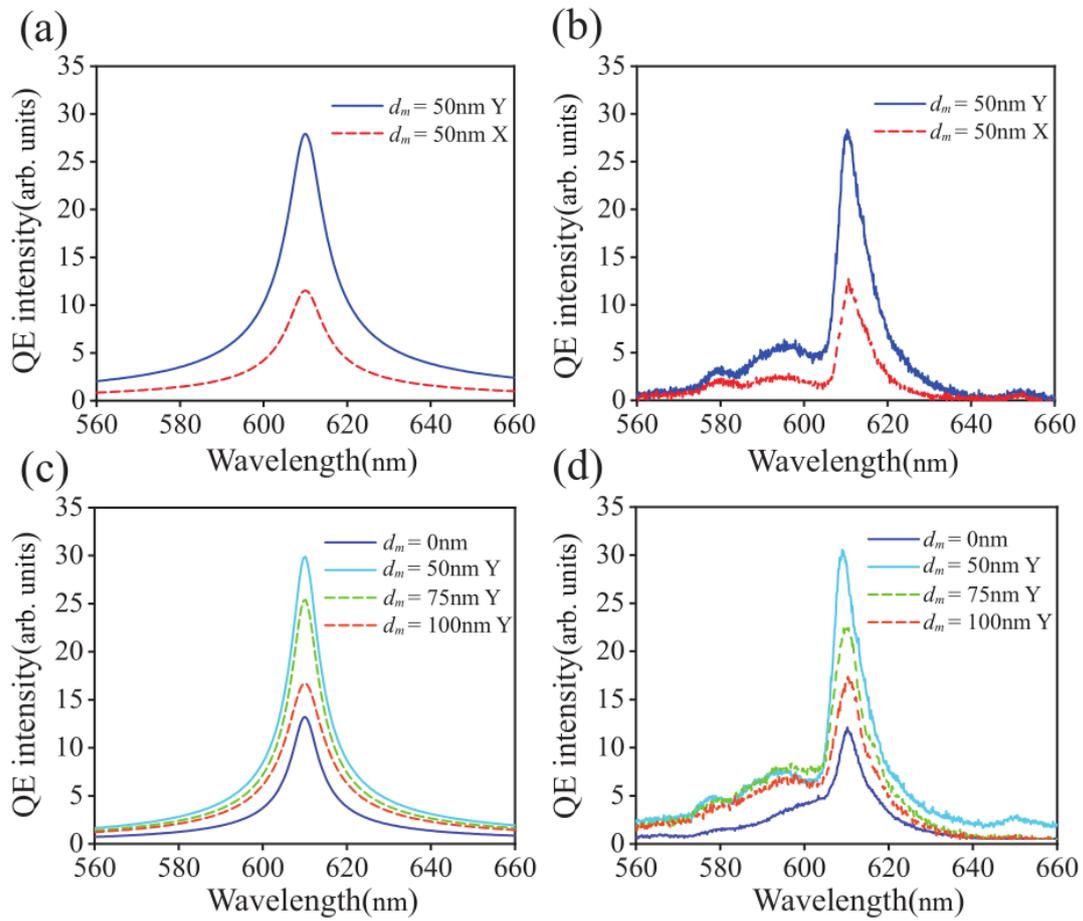

**Figure 6** (a) The QE intensity with different polarized states in the moiré lattice-metasurface structure with $d_m$=50nm. (b) The experimental measured QE intensity with different polarized states in the moiré lattice-metasurface structure with $d_m$=50nm. (c) The Y polarized QE intensity with different $d_m$ in the moiré lattice-metasurface structure. The pumping wavelengths correspond to the dots on the left side white region in Fig. 5(d). (d) The experimental measured Y polarized QE intensity with different $d_m$ in the moiré lattice-metasurface structure. The pumping wavelengths correspond to the dots on the left side white region in Fig. 5(f).